\documentclass[conference]{IEEEtran}
\IEEEoverridecommandlockouts
\usepackage{cite}
\usepackage{amsmath,amssymb,amsfonts}
\usepackage{multirow}
\usepackage{algorithm}  
\usepackage{algpseudocode}
\usepackage{graphicx}
\usepackage{textcomp}
\usepackage{xcolor}
\usepackage[mathlines,switch]{lineno}
\makeatletter
\let\old@ps@IEEEtitlepagestyle\ps@IEEEtitlepagestyle
\def\confheader#1
{\def\ps@IEEEtitlepagestyle
    {\old@ps@IEEEtitlepagestyle
     \def\@oddhead{\strut\hfill#1\hfill\strut}
     \def\@evenhead{\strut\hfill#1\hfill\strut}}
 \ps@headings}
\makeatother
\confheader
\def\BibTeX{{\rm B\kern-.05em{\sc i\kern-.025em b}\kern-.08em
    T\kern-.1667em\lower.7ex\hbox{E}\kern-.125emX}}

\begin{document}

\title{Towards Secure AI-driven Industrial Metaverse with NFT Digital Twins\\
\thanks{
This is an earlier version of a work accepted for publication at \textit{$17^{th}$ International Conference on COMmunication Systems \& NETworkS, 2025 (COMSNETS).}
}
}

\author{

\IEEEauthorblockN{Ravi Prakash, and Tony Thomas}
\IEEEauthorblockA{\textit{School of Computer Science and Engineering} \\
\textit{Kerala University of Digital Sciences, Innovation and Technology,}\\
Kerala, India \\
ravi.csres22@duk.ac.in, and tony.thomas@duk.ac.in}

}

\IEEEoverridecommandlockouts


\maketitle

\IEEEpubidadjcol

\begin{abstract}
The rise of the industrial metaverse has brought digital twins (DTs) to the forefront. Blockchain-powered non-fungible tokens (NFTs) offer a decentralized approach to creating and owning these cloneable DTs. However, the potential for unauthorized duplication, or counterfeiting, poses a significant threat to the security of NFT-DTs. Existing NFT clone detection methods often rely on static information like metadata and images, which can be easily manipulated. To address these limitations, we propose a novel deep-learning-based solution as a combination of an autoencoder and RNN-based classifier. This solution enables real-time pattern recognition to detect fake NFT-DTs. Additionally, we introduce the concept of dynamic metadata, providing a more reliable way to verify authenticity through AI-integrated smart contracts. By effectively identifying counterfeit DTs, our system contributes to strengthening the security of NFT-based assets in the metaverse.
\end{abstract}


\section{Introduction}
\label{sec:introduction}

The industrial metaverse (IMV) \cite{zheng2022industrial, lyu2023digital} is rapidly emerging as a paradigm-shifting technology that integrates the physical and digital worlds. Central to this paradigm are digital twins (DTs) \cite{bordegoni2023exploring, song2021build, han2022dynamic}, virtual representations of real-world assets or processes. DTs offer invaluable insights into asset performance, enabling predictive maintenance, optimization, and innovation. \\

Blockchain technology (BCT) \cite{prakash2022blockchain}, with its decentralized and immutable nature, has found numerous applications in the IMV \cite{song2021build}. Non-fungible tokens (NFTs) \cite{jing2022rarity} are a particular subset of blockchain assets that represent unique digital items \cite{gupta2023nft}. In the context of DTs, NFTs can be used to create and own cloneable DTs like the NFT avatars in social metaverse \cite{zelenyanszki2023privacy}, allowing for multiple virtual instances of a physical asset. This enables simultaneous testing and interaction across various virtual environments, accelerating development and experimentation. \\

However, the decentralized nature of NFTs also presents challenges due to possible security threats associated with BCT \cite{prakash2022blockchain}. The potential for unauthorized duplication, or counterfeiting, of NFT is a growing concern \cite{zelenyanszki2023privacy}. Existing methods for detecting fake clones often rely on static information like metadata, which can be easily manipulated. To address these limitations and ensure the integrity of NFT-DTs in the IMV, robust and reliable security measures are essential. 
This research focuses on developing a deep learning-based solution to identify counterfeit NFT digital twins. Inspired by behavioural biometrics \cite{pfeuffer2019behavioural, orgaz12clustering}, by analyzing the distinctive behavioural patterns of DTs, we aim to enhance the security and authenticity of these valuable digital assets in the industrial metaverse. \\

The remainder of this paper is organized as follows: Section \ref{sec:litrev} reviews existing literature on the industrial metaverse and its components. In Section \ref{sec:method}, we describe the architecture and operation of our proposed hybrid NFT-DT classification model. Subsequently, Section \ref{sec:exp_setup} outlines the experimental setup designed to simulate IMV scenarios for DTs of heaters. Section \ref{sec:security_analysis} discusses potential security threats associated with DTs in the metaverse. Finally, the results of our experiments are presented in Section \ref{sec:res}, leading to our conclusions and future research directions in Section \ref{sec:concNscope}.

\section{Literature Review} 
\label{sec:litrev}

\subsection{Industrial Metaverse} 

The metaverse \cite{park2022metaverse} is a complex and immersive environment that integrates several key components to create a fully interactive experience across various domains including healthcare \cite{hulsen2024applications, lee2023counseling}, education \cite{kaddoura2023rising, lepez2022metaverse}, power managements \cite{zhou2022metaverse}. It is often categorized in the social \cite{cheng2023metaverse, falchuk2018social} and industrial \cite{kuru2023metaomnicity, bordegoni2023exploring, zheng2022industrial} scenario. In \cite{prakash2023security}, authors presented the healthcare metaverse and its three essential components including Extended Reality (XR), Blockchain Technology (BCT), and user avatars. The work describes that three more dimensions add to these components due to the their combinations. However, these all components are essential only for a highly immersive metaverse scenarios like a critical medical application. Hence, the components proposed in \cite{prakash2023security} can be applicable in a generalized form of the metaverse, which can be used completely or partially based on the complexity and immersiveness of application scenario. \\

The industrial metaverse (IMV) is a form of the metaverse where physical and digital worlds are integrated virtually to serve the industrial solution like virtual power-grid \cite{deng2023metaverse, zhou2022metaverse}, interactive smart city model \cite{kuru2023metaomnicity, song2021build}, and industrial production line \cite{lyu2023digital, han2022dynamic}. Similar to avatars, that represents individuals in social context, the counter representation of devices and other interactive objects in virtual scenarios is referred to as the ``digital twin" (DT). Another critical component of IMV is blockchain technology (BCT), which is applicable in various aspects like data security, traceability, and verifying the ownership within the metaverse. The BCT, along with DT, enhances the delivery of services such as, warehouse monitoring, remote training, and supply chain traceability. The BCT introduce a new dimensions of engagement and accessibility, making the IMV a promising frontier across the domains. 

\subsection{NFTs in Metaverse} 

NFTs are the digital assets traceable on the blockchains. Every NFT holds a unique value which makes it different from the crypto currency that are fungible tokens. In \cite{gupta2023nft}, authors explored the transformative impact of NFTs on artists and the art market. Here authors highlighted the role of blockchain technology in enabling artists, game developers, and content creators to tokenize their work to reach a global audience and retain ownership. Although the work envisioned a future where NFTs play a crucial role in personalized digital interactions, yet, this work did not provide any insights into the application of NFTs in the metaverse. In another work by Jing et al. \cite{jing2022rarity}, the relationship between rarity and the aesthetic experience of 3D profile picture NFTs was studied in the metaverse scenario. The study primarily suggests that 3D profile picture NFTs play a significant role in the metaverse by serving as ``digital avatars" helping users to project themselves in an immersive virtual world. 
As avatar customization is possible in immersive environments, this study emphasizes the importance of understanding these dynamics for creating valuable and engaging NFT experiences. 
Musamih et al. \cite{musamih2022nfts} also presented a vision for NFTs, particularly in the healthcare domain, emphasizing their potential applications in patient data management, supply chain transparency, and clinical trial integrity. They highlighted the importance of patient empowerment through NFT ownership and the need for secure digital twins to represent patients and their data. Notably, here authors have considered tokenizing the medical assets through NFTs to produce the DTs. 
Another DT-based strategy was proposed in \cite{wang2023nft} to preserve and sustain Miao silver craftsmanship in the metaverse era using NFTs. This paper highlights the potential of using NFT DTs in fields like fashion, art, and cultural events. The work claimed that DTs of artifacts can provide verifiable information about their origin, history, and ownership by leveraging the principles of BCT. \\

The fake avatar cloning in  metaverse can lead to various security risks. We also came across a privacy awareness framework for NFT avatars in the metaverse where avatars were represented as NFTs \cite{zelenyanszki2023privacy}. This work focused on recognizing the cloned NFT-avatars based on some similarity score. Such cyber threat may also hold in IMV, however, we could not find any work done towards the security of DTs in IMV. 

\subsection{Behavioural Pattern in Metaverse}


Different combinations of body motion features can be analyzed to get the behavioural patterns \cite{pfeuffer2019behavioural}. The behavioural patterns, which are unique to each individual, are known as `behavioural biometric' and can be used for individual identification. 
In \cite{orgaz12clustering}, authors explore the use of clustering techniques to analyze avatar behaviours in virtual worlds. By extracting data on avatar movements, interactions, and chat logs, the researchers aim to automatically classify avatars into different groups based on their observed behaviours. In some other works in this domain, behavioural patterns are used to identify individuals \cite{nair2023unique}. 
However, it has been found that these are not limited to human behaviour as devices \cite{freire2019deep}, software \cite{khoshgoftaar1994comparative, zolkipli2011approach} and industrial control systems \cite{varghese2022digital} can also be identified by the behavioural patterns using various approaches including ML. These works suggest the fake components or devices can be identified by observing  anomalous or fraudulent behaviours. \\

In a  recent work by Zelenyanszki et al. \cite{zelenyanszki2023privacy}, they proposed a privacy awareness framework for NFT avatars in the metaverse. This work primarily focused on integrating avatars as NFTs and detecting cloned avatars by evaluating their similarity using a scoring system. They have also discussed the possibility of using machine learning algorithms to identify potential privacy threats and enhancing user awareness through behavioural tracking. However, the work lacks discussion on the essential behavioural patterns in details and validating it through any experiment. Taking into account these diverse scenarios and insights gained from work in the metaverse, we chose to jointly leverage blockchain technology (BCT) and machine learning to secure NFT digital twins within the IMV ecosystem.

\section{Methodology} 
\label{sec:method}

\begin{figure}[!ht]
    \centering
    \includegraphics[width=\linewidth]{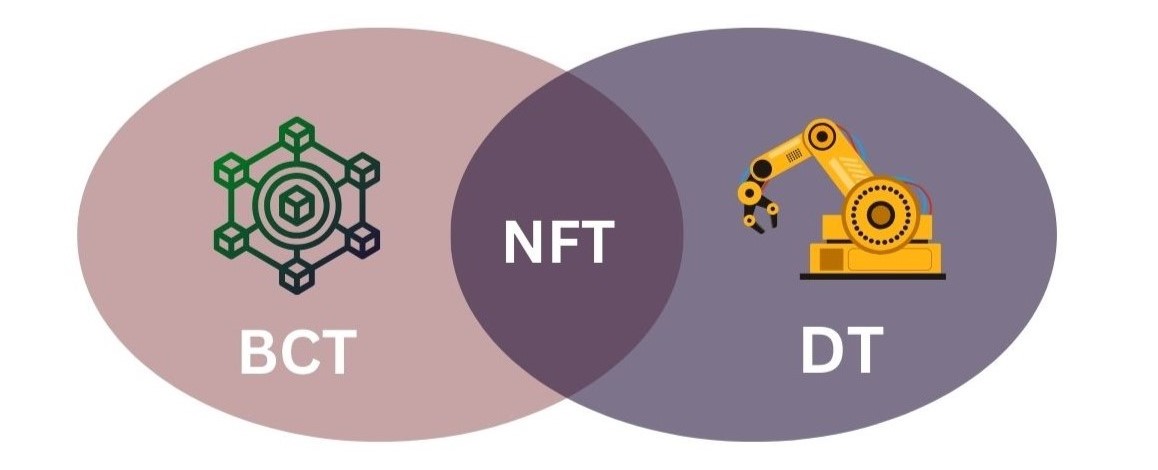}
    \caption{Components of Industrial Metaverse}
    \label{fig:dt_metaverse}
\end{figure}

\begin{figure*}[!htbp]
    \centering
    \includegraphics[scale=0.4]{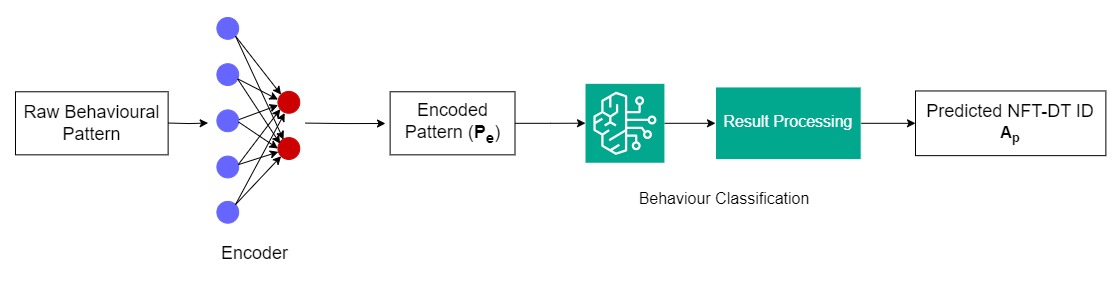}
    \caption{DT classification from raw behavioural patterns}
    \label{fig:Arch_DL_combined}
\end{figure*}

IMV consists of the devices and other physical entities in industrial applications. The three core components of an ideal industrial metaverse are - BCT, DT and NFT, as shown  in Fig. \ref{fig:dt_metaverse}. In this paper, we propose the development of a secure metaverse system for digital twins by introducing NFT digital twins (NFT-DTs) and AI-driven smart contracts. These smart contracts dynamically verify the originality of a digital twin based on its behavioural patterns. The complete DT behavioural pattern recognition is split into two parts - \textit{a) pattern encoding, b) classification}, both addressed using deep learning models (Fig. \ref{fig:Arch_DL_combined}). For hyperparameter optimization of these models, we employed the Python library \textit{keras-tuner}.

\subsection{Pattern Encoding}
\label{subsec:patt_enc_DAE}

Autoencoders are artificial neural networks (ANNs) designed to compress and reconstruct input data. These models consist of an encoder component that compresses the input into a latent representation as $P_e$ and a decoder that reconstructs the original input from this latent representation or encoding. \\

The sample of a raw behavioural pattern of a DT contains the $t$  readings of $k$ features. Hence, the behavioural pattern is a 2-dimensional (2-D) array of size $(t \times k)$. Hence, a denoising autoencoder (DAE) was designed to encode the 2-D patterns to a 1-D array of $k'$ elements. Conversely, the decoder aims to reconstruct the original behavioural pattern from the encoded representation $P_e$. The weights of the autoencoder are optimized in order to minimize the reconstruction loss of the complete system. The proposed solution utilizes an encoder to update the metadata of the NFT-DT with the most recent encoded patterns and enables real-time pattern recognition of the digital twin when integrated with a classifier. \\ 


The encoder of the proposed DAE initially transforms the 2-D time series pattern as a 1-D array of $tk$ elements. The transformed array is then subjected to a hierarchical encoding process using five hidden layers, progressively compressing the information into a 1-D array of $k'$ element vectors. Here, each of the hidden layers includes a normalization layer followed by a dropout, which drops the noise from the input patterns. The complete encoder model was dveloped for $t=34$, $k=5$, and $k'=62$. 


\subsection{Behavioural Pattern Classification}
\label{subsubsec:patt_clf_GRU}

Coming from the family of deep learning, RNN models can robustly handle complex sequential data \cite{keren2016convolutional, chung2015recurrent}. Equation  \ref{eq:hiddenSt_rnn} and Equation \ref{eq:opSt_rnn} provide the calculation for a hidden state $h_t$ and output $y_t$ at time $t$. Here, $x$ and $y$ represent the input and output values, $W$ denotes the weight matrix, and $b$ denotes the bias matrix. However, the traditional RNN models suffer from the vanishing gradient problem \cite{hochreiter1998vanishing}. Therefore, these models fail to retain the essential temporal information to learn from the long-term dependencies.

\begin{equation}
    h_t = tanh(W_{hh}h_{t-1} + W_{xy}x_t + b_h)
    \label{eq:hiddenSt_rnn}
\end{equation}
\begin{equation}
    y_t = tanh(W_{hy}h_{t} + b_h)
    \label{eq:opSt_rnn}
\end{equation}

\begin{figure}[htbp]
    \centering
    \includegraphics[width=\linewidth]{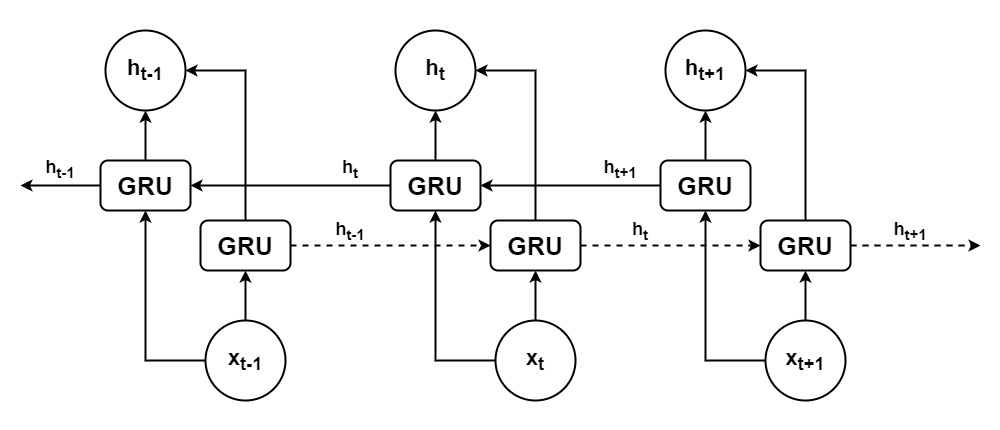}
    \caption{Data flow in a Bi-GRU}
    \label{fig:biGRU_3Stt}
\end{figure}

To overcome this problem, the gated recurrent unit (GRU), another type of RNN, can be used that is specialized to learn from the long-term dependencies \cite{chung2014empirical, dey2017gate}. GRU aim to update the hidden states with some additional values through the reset gates $r_t$ and update gate $z_t$, at the time $t$, as per the Eqn. \ref{eq:resetSt_gru}, Eqn. \ref{eq:updSt_gru}, and Eqn. \ref{eq:updSt_gru}, respectively. Additionally, the order sequence also matters in time-series data and can be used to generate or identify the past or future sequences. Therefore learning from both past and future insights of a time-series input can lead to a more robust model. An advanced version of GRU, bidirectional GRU (Bi-GRU), is capable of learning such insights by processing the input in both orders (Fig. \ref{fig:biGRU_3Stt}). Hence, we developed a pattern classification model based on the Bi-GRU.

\begin{equation}
    r_t = \sigma(W_{xr}x_t + W_{hr}h_{t-1} + b_r)
    \label{eq:resetSt_gru}
\end{equation}
\begin{equation}
    z_t = \sigma(W_{xz}x_t + W_{hz}h_{t-1} + b_z)
    \label{eq:updSt_gru}
\end{equation}
\begin{equation}
    h_t = (1 - z_t) \circ h_{t-1} + z_t \circ tanh(W_{xh}x_t + r_t \circ W_{hh}h_{t-1} + b_h)
    \label{eq:hiddenSt_gru}
\end{equation}


\begin{figure}[htbp]
    \centering
    \includegraphics[width=0.7\linewidth]{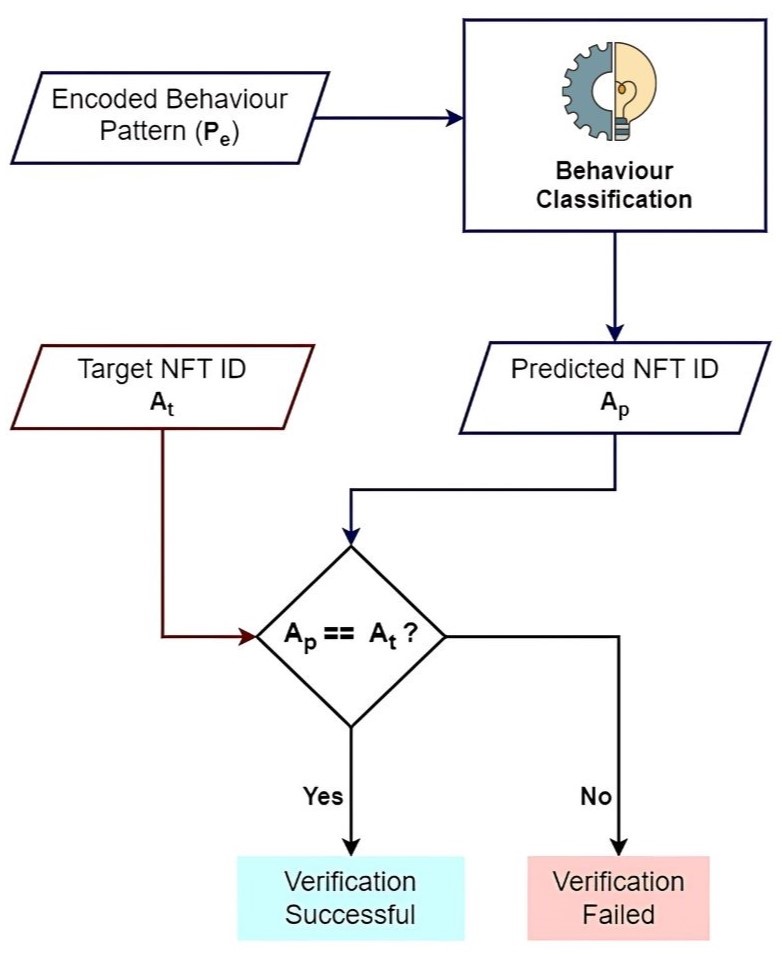}
    \caption{DT verification based on the pattern classification}
    \label{fig:pattern_classification}
\end{figure}

Although, the encoded patterns $P_e$ do not explicitly have any separation of each time-stamp, however, this information is efficiently compressed in the encoding. Hence, the proposed classification model works with the encoded data patterns provided by the encoder component of DAE, discussed in Section \ref{subsec:patt_enc_DAE}. The final classification results are later processed to identify the in-class and OOC behavioural patterns. The working model of the proposed classification system is shown in Fig. \ref{fig:pattern_classification}. \\

We developed a multi-class classification model for DT behavioural pattern classification using Bi-GRU. This approach enhances the DT behavioural pattern classification in the following ways:

\begin{enumerate}
    \item The original behavioural pattern is protected as only encoded information is supplied to the classifier, which may be publicly accessible as an open DT detection solution.
    \item Real-time and faster classification due to the smaller size of the input.
\end{enumerate}

\section{Experimental Setup}
\label{sec:exp_setup}
The testbed was built using a local Ethereum test network utilizing Ganache as the experimental environment. For better usage, a command-line interface (CLI) application, developed with Hardhat and JavaScript, facilitated interaction with the test network nodes. We also created separate scripts, written using \textit{ether.js} library, to query, compare, and engage with NFT-DTs accessible on the network. The Solidity programming language was used to construct smart contracts, leveraging the OpenZeppelin Contracts library and ERC721 interface. The static metadata for these NFT-DTs was hosted on a separate GitHub repository. 
This testbed helped to mint the NFT-DTs and create their clones and supports the DT creation in the following ways:

\begin{enumerate}
    \item Minting with uniform resource identifier (URI) of metadata \textit{(genuine way of fresh DT creation)}; 
    \item Cloning using URI; 
    \item Cloning using token ID of DT, provided by smart contract; 
    \item Cloning using metadata populated randomly. \\ 
\end{enumerate}

\begin{figure}[htbp]
    \centering
    \includegraphics[width=\linewidth]{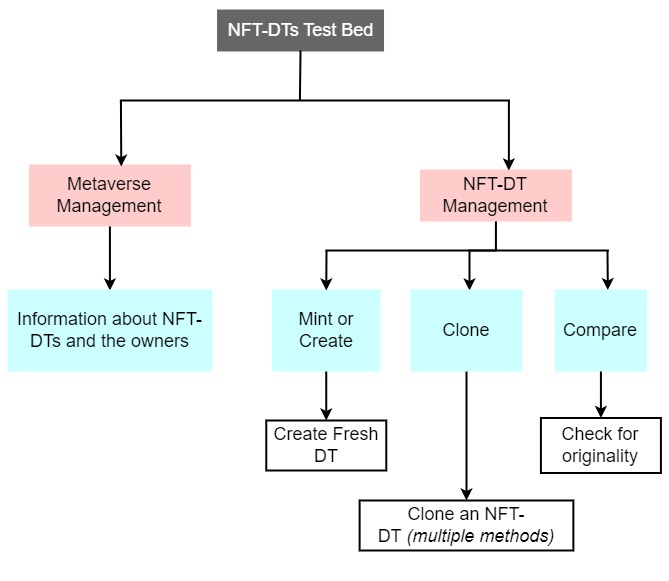}
    \caption{Features of the test bed}
    \label{fig:testBed_features}
\end{figure}

The DTs were minted, cloned, and analyzed for the security of existing clone detection methods. We have also proposed the concept of dynamic metadata for the NFT-DTs for enhanced security, and it can be integrated with the current test bed. 
The utilities provided by our test bed for experiments are shown in Fig. \ref{fig:testBed_features}. Here, DT comparison functionality is utilized to differentiate between the cloned and original NFT-DTs.

\subsection{Experimental Set-up for DT Behavioural Pattern Recognition}
\label{subsec:behPattRecog_avatar}

We consider the scenario of the industrial metaverse where the digital twin of heaters is to be used as NFT-DTs. As discussed in Section \ref{sec:method}, the proposed method of NFT comparison (Fig. \ref{fig:Arch_DL_combined}), based on real-time behavioural patterns, is implemented. \\


\subsubsection{Dataset}  

We used a public dataset from Kaggle related to the health of the digital twins of gadgets \cite{HeaterHel_Data}. It consists of the readings collected for heaters of the following elements over a continuous interval of time:

\begin{enumerate}
    \item \textit{Voltage measured:} The voltage supplied 
    \item \textit{Current measured:} The current supplied 
    \item \textit{Temperature measured:} The temperature measured 
    \item \textit{Current charge:} The current flow recorded while charging 
    \item \textit{Voltage charge:} The voltage level while charging 
    \item \textit{Capacity:} The energy storage capacity \\ 
\end{enumerate}

After the pre-processing, the \textit{Voltage measured} and \textit{Voltage charge} were found to be identical. Hence, \textit{Voltage charge} was excluded from further usage. In various cycles, the data is recorded for $N=4$ \textit{heaters}. Here, the samples collected during each cycle of the experiment also varied. We considered the 132 cycles of experiments where the duration for each cycle ranged from 26 minutes to 60 minutes. The complete information is provided in the Table \ref{tab:dt_data}. The proposed classification model is meant to be accessible to each of the nodes in the blockchain network of the metaverse. This enables all nodes to individually verify the originality of an NFT-DT by passing the latest behaviour patterns.

\begin{table}[htbp]
    \centering
    \caption{Data related to digital twin of heaters used}
    \begin{tabular}{|l|c|}
        \hline
        \textbf{Property} & \textbf{Measurement} \\
        \hline 
        Total DT     &   4 \\
        \hline
        Experiment Cycles / DT  &   132 \\
        \hline
        Data Reading / Cycles   &   170 \\
        \hline
        Rows / Input Sample     &   34 \\
        \hline
        Features / Input        &   5 \\
        \hline
        Training Samples        &   2112 \\
        \hline
        Test Samples            &   528 \\
        \hline
        Total Data Samples      &   2640 \\
        \hline 
    \end{tabular}
    \label{tab:dt_data}
\end{table}

\section{Security Analysis}
\label{sec:security_analysis}

Various attacks related to the enabling technologies of the metaverse have also been discovered which can affect the metaverse in terms of data loss, unauthorized access, and denial of services (DoS), to name a few \cite{prakash2023security}. 
In this section, we have analyzed multiple security aspects related to DTs in the metaverse, the limitations of existing solutions, and the proposed approach for overcoming the problems. 

\subsection{Issues with DT Cloning}
\label{subsec:avtClon_issue}

The principles of BCT can be largely helpful in addressing the problems related to information traceability and verifiability of DTs using NFT-DTs. However, blockchain has also been found to be vulnerable to several attacks. The vulnerabilities associated with smart contracts, and targeted market risks can also affect the NFTs \cite{prakash2022blockchain}, and hence NFT-DTs. Apart from these possible dimensions of risks associated with the NFT-DTs, the possibility of DT cloning can also lead to severe problems. However, genuine DT cloning can be helpful in many ways \cite{fernando2005cloning}, but it may also lead to the fake cloning of NFT-DTs. \\

Table \ref{tab:geVSfake_clones} lists some common differences between a genuine and fake clone of DTs in socio-industrial metaverse scenarios. The clones of DTs can enable user presence in multiple virtual environments and can be used for testing purposes, yet a fake clone can still lead to information leakage and account takeover. Authors of \cite{zelenyanszki2023privacy} have addressed this issue mostly in the case of the social metaverse by leveraging the comparison of metadata and the appearance of two avatars in a metaverse environment. They have also conceptualized the behavioural comparisons of human avatars, however, their main work highlights how static metadata and image distance comparison can be utilized to detect cloned NFTs. \\

\begin{table}[htbp]
    \centering
    \caption{Difference between the genuine and fake DTs clones}
    \begin{tabular}{|p{4cm}|p{4cm}|}
        \hline
        \textbf{Genuine Clone} & \textbf{Fake Clone} \\
        \hline
        Have identical metadata & Have dissimilar Metadata \\
        \hline
        Identical URI & Different URI \\
        \hline
        Similar behavioural patterns & Dissimilar behavioural patterns \\
        \hline
        Identical source for DTs &  Different sources for DTs \\
        \hline
    \end{tabular}
    \label{tab:geVSfake_clones}
\end{table}

\subsection{Limitations of existing DT clone detection method}
\label{subsec:avtDetect_issue}

Existing NFT clone detection methods utilize the comparison of static information to find the similarity between original and target NFTs \cite{zelenyanszki2023privacy}. This is why, the duplication of static information like metadata, URL of additional data locations and other information associated with NFT can lead to the highly similar fake clones of NFT. Our experiments prove that there is a high chance of creating a fake NFT clone, making it difficult to detect the fraudulent cloning of NFT-DTs. The details of this experiment are provided in Section \ref{sec:res}. 

\subsection{Securing the Behavioural Patterns}
\label{subsec:behPatt_sec}

Due to the uniqueness of patterns \cite{freire2019deep, khoshgoftaar1994comparative, zolkipli2011approach, varghese2022digital}, we decided to utilize the behavioural patterns of DTs for originality verification. However, as the pattern itself is sensitive information, we ensured to protect the same by encoding them using DAE. Automating the process of pattern encoding through smart contracts and only considering the encoded patterns throughout the verification process would lead to a more secure metaverse application. \\

In the context of a multi-class classifier, completeness is the measurement of not identifying an input as a class which does not exist. Therefore, a complete solution is resistant to targeted attacks like identity spoofing. We tested $132$ behavioural patterns for each of the $4$ different NFT-DTs and computed the completeness scores for each DT. The final results are provided in Table \ref{tab:completeness_score}.

\begin{table}[htbp]
    \centering
    \caption{Completeness score for $4$ NFT-DTs}
    \begin{tabular}{|c|c|}
        \hline
        \textbf{NFT-DT} & \textbf{Completeness} \\
        \hline
        DT-$0$     &       $100.00\%$ \\
        \hline
        DT-$1$     &       $97.73\%$ \\
        \hline
        DT-$2$     &       $100.00\%$ \\
        \hline
        DT-$3$     &       $100.00\%$ \\
        \hline
    \end{tabular}
    \label{tab:completeness_score}
\end{table}

The most severe attack against the proposed solution would aim the DT behavioural pattern spoofing to succeed in account takeover. However, Table \ref{tab:completeness_score} indicates that the model is robust enough to address such attacks, as most of the DTs have a completeness score of $100.00\%$. The minimum completeness score, corresponding to DT ID-$1$, is $97.73\%$, which indicates that the maximum $6$ (less than $3\%$) of the $396$ samples were misclassified as an NFT-DT with ID-$1$. Hence, this experiment tells that out of every $528$ attempt, the attacker could only succeed $6$ ($1.14\%$) times to spoof the target. Although the completeness remained constant even after setting the $\tau$ for OOC rejection, we noticed a nominal downfall in the case of soundness. The additional details related to the soundness and completeness score are given in Section  \ref{sec:res}.

\subsection{Dynamic metadata for NFT-DTs}
\label{subsec:dynMetaData}

Each NFT is associated with a JSON metadata file containing additional information. The $attributes$ field within this metadata stores various properties related to the NFT's interactive and non-interactive characteristics. To strengthen DT security through real-time verification, we propose a new approach: dividing the $attributes$ field into two distinct fields: $readOnly\_attributes$ and $writable\_attributes$. This dynamic metadata concept allows for more robust security measures. During the minting process, an \textit{encoder} model ($\epsilon_{i}$) is also deployed, where $i$ denotes the ID of the current NFT-DT. The address of $\epsilon_{i}$ is stored along with the node information. The smart contract is configured to update the $readOnly\_attributes$ of metadata over a fixed interval of time as per the Algorithm \ref{algo:dynamic_metadata_update}. The smart contract allows a user to verify the originality of any NFT-DT by passing a behavioural pattern from cached data as $\beta^{c}_{i}$. Algorithm \ref{algo:nft_dt_verification} is triggered during the verification process of DT $i$, which can detect a genuine or fake clone of any NFT-DT in the network of IMV. \\


\begin{algorithm}[ht]
\caption{Dynamic Metadata Update for NFT-DTs}
\label{algo:dynamic_metadata_update}
\begin{algorithmic}[1]
\Require $\beta^{l}_{i}$: Live pattern of NFT-DT $i$
\Require $\epsilon_{i}$: Encoder model associated with NFT-DT $i$
\Require $readOnly\_attributes$ 

\State \textbf{Compress Live Pattern:}
\State $\beta^{l'}_{i} \gets \epsilon_{i}(\beta^{l}_{i})$ \Comment{Compress live pattern using encoder}

\State \textbf{Update Read-Only Attributes:}
\State Update $readOnly\_attributes$ with $\beta^{l'}_{i}$

\end{algorithmic}
\end{algorithm}


\begin{algorithm}
\caption{NFT-DT Verification}
\label{algo:nft_dt_verification}
\begin{algorithmic}[1]
\Require $\beta^{c}_{i}$: Cached behavioural pattern of NFT-DT $i$
\Require $\epsilon_{i}$: Encryption model associated with NFT-DT $i$
\Require $\beta^{l'}_{i}$: Latest behavioural pattern of NFT-DT $i$
\Require $C_{\alpha}$: Shared DT classification model

\State \textbf{Compress Cached Pattern:}
\State $\beta^{c'}_{i} \gets \epsilon_{i}(\beta^{c}_{i})$

\State \textbf{Load Latest Pattern from Metadata:}
\State Load $\beta^{l'}_{i}$ from metadata

\State \textbf{Perform Predictions:}
\State $P_{\alpha} \gets C_{\alpha}(\beta^{l'}_{i})$
\State $P_{c} \gets C_{\alpha}(\beta^{c'}_{i})$

\State \textbf{Verify Originality:}
\If{$P_{\alpha} \neq P_{c}$}
    \State \textbf{Declare Fake Clone:}
    \State \textbf{Print} ``Verification Failed \textit{(fake clone)}"
\ElsIf{$N$ is odd}
    \State \textbf{Print} ``Verification Successful \textit{(genuine clone)}"
\EndIf
\end{algorithmic}
\end{algorithm}

In the current scenario, we assume that the classifier ($C_\alpha$) is pre-trained and already deployed. The model may either be stored in the test network or outside the same but, the address of $C_\alpha$ must be known to each of the nodes. 
We have tested the performance of our clone detection model separately, which can be integrated with the metaverse for AI-driven smart contracts. \\

As mentioned in Table \ref{tab:geVSfake_clones}, the fake DT cloning can be done by copying the complete metadata and hosting it to a separate location (URI). The different URI enables the attacker to own the fake clone of NFT-DT and have complete control over it. With a static metadata, the comparison will always results into a genuine clone. However, due to the proposed concept of dynamic metadata through AI-driven smart contracts, this NFT-DT cloning attack can be mitigated. We can represent the dynamic metadata $\Delta M$ of an original DT as $\Delta M_o$, genuine clone of DT as $\Delta M_g$, and fake clone DT as $\Delta M_f$. Considering that $\Delta M^{\alpha_i}_{o_t}$ representes the metadata of an original NFT-DT $\alpha_i$, at time $t$, Eq. \ref{eq:md_atT} shows the metadata of 2 clones created at the same time $t$. 

\begin{equation}
    \begin{array}{ll}
    \Delta M^{\alpha_i}_{g_t} = \Delta M^{\alpha_i}_{o_t} \\
    \Delta M^{\alpha_i}_{f_t} = \Delta M^{\alpha_i}_{o_t}
    \end{array}
    \label{eq:md_atT}
\end{equation}

It can be noticed that both genuine ($\Delta M^{\alpha_i}_{g_t}$) and fake ($\Delta M^{\alpha_i}_{f_t}$) clones of an NFT-DT $\alpha_i$ at time $t$ are identical. The metadata updates are triggered by the temper-proof smart contract, which are meant to be separate for each NFT. In the given scenario, $\Delta M^{\alpha_i}_{g_t}$ is mapped to retrieve the data from already existing DT $\alpha_i$. However, for the fake NFT, attacker would try to map a new NFT-DT by mirroring the metadata. Hence, at a later time $t'$, the two metadata will have signifient difference, that can be expressed through Eqn. \ref{eq:md_atTdash}. The change occurs due to the difference in behavioural pattern of genuine and fake clones at time $t'$. Additionally, the manual tempering of $\Delta M^{\alpha_i}_{f_{t'}}$ is not possible during verification, due to the involvement of smart contract. Hence, the proposed $\Delta M$ for NFT-DTs strengthens the security and makes the the system resilient to the fake cloning using metadata. 

\begin{equation}
    \begin{array}{ll}
    \Delta M^{\alpha_i}_{g_{t'}} = \Delta M^{\alpha_i}_{o_{t'}} \\
    \Delta M^{\alpha_i}_{f_{t'}} \neq \Delta M^{\alpha_i}_{o_{t'}}
    \end{array} 
    \label{eq:md_atTdash}
\end{equation}

\vspace{0.1cm}

\section{Results and Discussion} 
\label{sec:res}

As provided in Table \ref{tab:dt_data}, the 34 continuous data rows are considered as one sample. Each of the samples has nearly 10 minutes duration. We proposed a hybrid classification system that utilizes an autoencoder and a GRU-based deep learning model. The proposed denoising auto-encoder takes an input matrix of size $34\times5$ and encodes it into an array of $62$ elements. The encoded data is passed through the classifier which is ultimately responsible for the identification of the correct digital twin based on the behavioural pattern of the device. \\

Using the testbed explained in Section \ref{sec:exp_setup}, we initially created $9$ original NFT-DTs, followed by creating $25$ clones based on the various methods. Table \ref{tab:dtCloneSim} presents the similarity score of some NFT-DT clones along with the corresponding target token IDs of NFTs. It was found that the similarity scores of cloned NFT-DTs ranged from $8.33\%$ to $100\%$. It indicated that there is a high chance of creating a fake NFT clone with the least similarity score. Hence, we provide a detailed performance of the proposed autoencoder for real-time behavioural pattern encoding and the proposed pattern-based NFT-DT classification system. A confidence-based threshold rule is also discussed which can help to identify the fake clones of a genuine DT. 

\begin{table}[htbp]
\caption{NFT-DTs and their clones}
\begin{center}
\begin{tabular}{|c|c|c|}
\hline
\textbf{Token ID}& \textbf{Clones} & \textbf{Metadata Similarity (\%)}\\
\hline

\multirow{2}{*}{1}  & Clone \#16 & 100\% \\
\cline{2-3}
& Clone \#33	& 91.67\% \\
\hline

\multirow{2}{*}{2} &	Clone \#13 &	8.33\% \\
\cline{2-3}
 &	Clone \#18 &	100\% \\
\hline

\multirow{3}{*}{3} &	Clone \#20 &	9.09\% \\
\cline{2-3}
 &	Clone \#30 &	63.64\% \\
\cline{2-3}
 &	Clone \#31 &	63.64\% \\
\hline

4	& Clone \#22 &	72.73\% \\
\hline

\multirow{2}{*}{8} &	Clone \#30 &	18.18\% \\
\cline{2-3}
 &	Clone \#34 &	25\% \\
\hline

\multirow{4}{*}{12} &	Clone \#27 &	100\% \\
\cline{2-3}
 &	Clone \#28 &	100\% \\
\cline{2-3}
 &	Clone \#32 &	58.34\% \\
\cline{2-3}
 &	Clone \#34 &	66.67\% \\
\hline

\end{tabular}
\label{tab:dtCloneSim}
\end{center}
\end{table}


We propose a real-time pattern analysis for a more efficient recognition of the originality of an DT. We evaluate our solution for DTs in industrial metaverse scenarios, however, a similar approach could be followed with avatars in the social metaverse. This process contributes in two major aspects - \textit{a) compressing large behavioural patterns for computational efficiency, b) real-time classification of an original and fake clone of NFT-DT}. \\

We developed an originality detection method for digital twins in an industrial metaverse context, for a heaters network. We considered a total of $4$ heaters for our experiments. The detection system was developed in 2 parts, \textit{a) data encoding}, and \textit{b) DT classification} using a denoising autoencoder (DAE) and bi-directional GRU-based classification model respectively, as discussed in Section \ref{sec:method}. The autoencoders aim to encode the $34$ continuous data recorded for $5$ sensors, to an array of $62$ numbers. the encoded data is then used for classification. \\

\subsection{Behavioural pattern analysis using DAE}

The proposed DAE compresses the raw behavioural patterns as an encoding of size $62$. The hyperparameter tuning was done using the \textit{keras-tuner} module in Python. The summary of model performance is provided in Table \ref{tab:dae_res}.

\begin{table}[htbp]
    \centering
    \caption{Performance of proposed DAE}
    \begin{tabular}{|l|c|}
        \hline
        \textbf{Properties} & \textbf{Measurements} \\
        \hline 
        Input data Size     &   ($34 \times 5$) \\
        \hline
        Output data Size  &   $(62, )$ \\
        \hline
        Mean Squared Error   &   $0.083$ \\
        \hline 
    \end{tabular}
    \label{tab:dae_res}
\end{table}

\subsection{Originality check for an NFT-DT}

The data was collected from $34$ continuous rows of behavioural patterns which can be considered for temporal feature analysis, where the order has significant importance. It is clear from the performance that it precisely compresses the sequential and temporal information within $62$ character. Hence, we developed a bi-directional GRU-based data classification method that works with the encoded data only. The performance of the proposed classifier is given in Table \ref{tab:clf_res}.  \\

\begin{table}[htbp]
    \centering
    \caption{Performance of classifier on encoded data}
    \begin{tabular}{|l|c|}
        \hline
        \textbf{Properties} & \textbf{Measurements} \\
        \hline 
        Input data Size     &   $62$ \\
        \hline
        Output data Size  &   $4$ \\
        \hline
        Loss   &   $0.058$ \\
        \hline
        Accuracy   &   $98.30\%$ \\
        \hline
        TP  &   519 \\
        \hline
        FP  &    9 \\
        \hline 
    \end{tabular}
    \label{tab:clf_res}
\end{table}

\begin{figure}[ht]
    \centering
    \includegraphics[width=\linewidth]{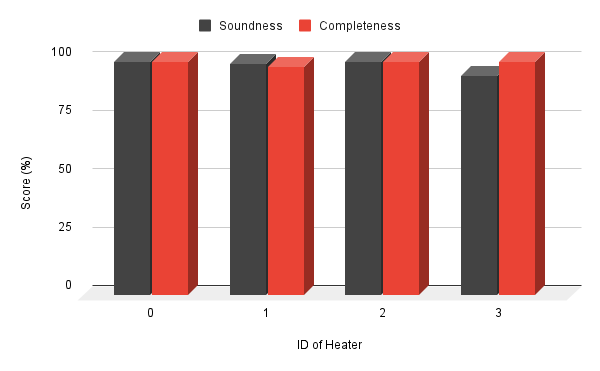}
    \caption{Soundness and Completeness score of proposed solution for 4 NFT-DT}
    \label{fig:sNc_score}
\end{figure}

Among $528$ test samples, the proposed model could not identify the $9$ samples. An in-depth analysis indicated that out of the $9$ misclassification, $8$ samples belonged to a DT with \textit{PhID} 3 and $1$ belonged to \textit{PhID} 1. Additionally, these are misclassified to DT with \textit{PhID} 1 and 3 respectively. However, among $132$ samples for each digital twin, a maximum $8$ misclassification brings the worst-case accuracy of an individual (soundness) DT to $93.94\%$ which might be considered satisfactory. The graph corresponding to the soundness and completeness of the solution is provided as Fig. \ref{fig:sNc_score}.  \\

\begin{figure}[!h]
    \centering
    \includegraphics[width=\linewidth]{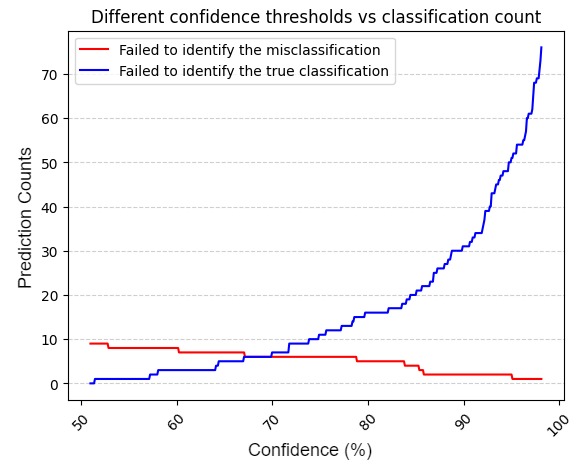}
    \caption{Optimal threshold $\tau$ was found to be $69.50\%$}
    \label{fig:tau}
\end{figure}

However, one major problem with such a multiclass classification model is that it can not accurately identify the pattern from an OOC/unseen DT. Hence, the model currently identifies the OOC DT as one of the $N$ known classes. 
To address the challenge of identifying OOC behaviours, we propose a novel approach based on prediction confidence. By analyzing the confidence levels associated with true and false predictions, we observed a clear discrepancy. Specifically, true predictions relatively exhibited higher confidence scores compared to false ones. Therefore, we introduced a threshold parameter, $\tau$, to effectively detect OOC DT instances. As per the rule, if the prediction confidence falls below $\tau$, it is considered to be the behaviour of an OOC DT. Fig. \ref{fig:tau} shows the results of our experiments to optimize the value of $\tau$, balancing the trade-off between true and false predictions. Through the analysis of true and misclassified samples, we determined the optimal threshold value to be $\tau=69.50\%$. \\

The final results obtained after setting the $\tau$ are provided in Table \ref{tab:clf_res_withTau}. Here \textit{false positive} (FP) represents the misclassified samples with a confidence higher than $\tau$ and \textit{false negative} (FN) represents the count of correct classifications, which are rejected due to confidence lower than $\tau$. \\

\begin{table}[htbp]
    \centering
    \caption{Results after the optimal value of $\tau$}
    \begin{tabular}{|l|c|}
        \hline
        \textbf{Evaluation Matrix} & \textbf{Counts} \\
        \hline 
        Total Samples     &   $528$ \\
        \hline
        $\tau$  &   $69.50\%$ \\
        \hline
        Accuracy     &   $97.73\%$ \\
        \hline
        FRR     &       $1.72\%$ \\
        \hline
        TP  &   $513$ \\
        \hline
        TN   &   $3$ \\
        \hline
        FP   &   $6$ \\
        \hline
        FN  &   $6$ \\
        \hline
    \end{tabular}
    \label{tab:clf_res_withTau}
\end{table}


\section{Conclusion and Future Scope}
\label{sec:concNscope}

This paper explored the combined role of NFT digital twins (DTs) and deep learning in shaping a more secure industrial metaverse. The NFT-DTs can enhance security using the BCT principles and ease of interaction through cloning. However, NFT-DT cloning can lead to security risks of counterfeit products and twin identity spoofing. We analyzed the data privacy and security associated with digital twins (DTs) and the performance of the existing NFT clone detection methods, in the context of the industrial metaverse. We then proposed a novel approach for clone originality verification through a hybrid deep-learning framework based on DAE and Bi-GRU, and mitigating the risk of fake cloning through dynamic metadata ($\Delta M$). The solution is developed for real-time behavioural pattern recognition and can be utilized to create AI-driven smart contracts. We also proposed the concept of dynamic metadata for NFT avatars making it difficult for attackers to clone the avatars using metadata. \\

Despite this paper introducing the feasibility of NFT-DTs for the secured industrial metaverse, the performance of DT clone detection with the behavioural classification of NFT-DTs can further be improved with sufficient OOC data to get more realistic results. Additionally, the multiclass classifications model can be difficult to deploy into an actual industrial metaverse due to the higher dependency on the behavioural data from all DTs. Hence, developing suitable one-class classifiers or federated learning-based approaches should also be explored to determine the feasibility of the proposed solution in a wider scenario.


\bibliographystyle{IEEEtran}
\bibliography{references}

\begin{thebibliography}{10}
\providecommand{\url}[1]{#1}
\csname url@samestyle\endcsname
\providecommand{\newblock}{\relax}
\providecommand{\bibinfo}[2]{#2}
\providecommand{\BIBentrySTDinterwordspacing}{\spaceskip=0pt\relax}
\providecommand{\BIBentryALTinterwordstretchfactor}{4}
\providecommand{\BIBentryALTinterwordspacing}{\spaceskip=\fontdimen2\font plus
\BIBentryALTinterwordstretchfactor\fontdimen3\font minus \fontdimen4\font\relax}
\providecommand{\BIBforeignlanguage}[2]{{%
\expandafter\ifx\csname l@#1\endcsname\relax
\typeout{** WARNING: IEEEtran.bst: No hyphenation pattern has been}%
\typeout{** loaded for the language `#1'. Using the pattern for}%
\typeout{** the default language instead.}%
\else
\language=\csname l@#1\endcsname
\fi
#2}}
\providecommand{\BIBdecl}{\relax}
\BIBdecl

\bibitem{zheng2022industrial}
Z.~Zheng, T.~Li, B.~Li, X.~Chai, W.~Song, N.~Chen, Y.~Zhou, Y.~Lin, and R.~Li, ``Industrial metaverse: Connotation, features, technologies, applications and challenges,'' in \emph{Asian Simulation Conference}.\hskip 1em plus 0.5em minus 0.4em\relax Springer, 2022, pp. 239--263.

\bibitem{lyu2023digital}
Z.~Lyu and M.~Fridenfalk, ``Digital twins for building industrial metaverse,'' \emph{Journal of Advanced Research}, 2023.

\bibitem{bordegoni2023exploring}
M.~Bordegoni and F.~Ferrise, ``Exploring the intersection of metaverse, digital twins, and artificial intelligence in training and maintenance,'' \emph{Journal of Computing and Information Science in Engineering}, vol.~23, no.~6, p. 060806, 2023.

\bibitem{song2021build}
Y.~Song and S.~Hong, ``Build a secure smart city by using blockchain and digital twin,'' \emph{International Journal of Advanced Science and Convergence}, vol.~3, no.~3, pp. 9--13, 2021.

\bibitem{han2022dynamic}
Y.~Han, D.~Niyato, C.~Leung, D.~I. Kim, K.~Zhu, S.~Feng, X.~Shen, and C.~Miao, ``A dynamic hierarchical framework for iot-assisted digital twin synchronization in the metaverse,'' \emph{IEEE Internet of Things Journal}, vol.~10, no.~1, pp. 268--284, 2022.

\bibitem{prakash2022blockchain}
R.~Prakash, V.~Anoop, and S.~Asharaf, ``Blockchain technology for cybersecurity: A text mining literature analysis,'' \emph{International Journal of Information Management Data Insights}, vol.~2, no.~2, p. 100112, 2022.

\bibitem{jing2022rarity}
T.~Jing and L.~Wang, ``How the rarity influences aesthetic experience of 3d profile picture nfts,'' in \emph{Proceedings of the Tenth International Symposium of Chinese CHI}, 2022, pp. 151--160.

\bibitem{gupta2023nft}
M.~Gupta, D.~Gupta, and A.~Duggal, ``Nft culture: A new era,'' \emph{Scientific Journal of Metaverse and Blockchain Technologies}, vol.~1, no.~1, pp. 57--62, 2023.

\bibitem{zelenyanszki2023privacy}
D.~Zelenyanszki, Z.~H{\'o}u, K.~Biswas, and V.~Muthukkumarasamy, ``A privacy awareness framework for nft avatars in the metaverse,'' in \emph{2023 International Conference on Computing, Networking and Communications (ICNC)}.\hskip 1em plus 0.5em minus 0.4em\relax IEEE, 2023, pp. 431--435.

\bibitem{pfeuffer2019behavioural}
K.~Pfeuffer, M.~J. Geiger, S.~Prange, L.~Mecke, D.~Buschek, and F.~Alt, ``Behavioural biometrics in {VR}: Identifying people from body motion and relations in virtual reality,'' in \emph{Proceedings of the 2019 CHI Conference on Human Factors in Computing Systems}, 2019, pp. 1--12.

\bibitem{orgaz12clustering}
G.~Orgaz \emph{et~al.}, ``Clustering avatars behavioursfrom virtual worlds interactions,'' in \emph{International Workshop On Web Intelligence \& Communities-WI\&C}, vol.~12, no.~4, 2012, pp. 1--7.

\bibitem{park2022metaverse}
S.-M. Park and Y.-G. Kim, ``A metaverse: Taxonomy, components, applications, and open challenges,'' \emph{IEEE access}, vol.~10, pp. 4209--4251, 2022.

\bibitem{hulsen2024applications}
T.~Hulsen, ``Applications of the metaverse in medicine and healthcare,'' \emph{Advances in Laboratory Medicine/Avances en Medicina de Laboratorio}, vol.~5, no.~2, pp. 159--165, 2024.

\bibitem{lee2023counseling}
K.~Lee, ``Counseling psychological understanding and considerations of the metaverse: A theoretical review,'' in \emph{Healthcare}, vol.~11, no.~18.\hskip 1em plus 0.5em minus 0.4em\relax MDPI, 2023, p. 2490.

\bibitem{kaddoura2023rising}
S.~Kaddoura and F.~Al~Husseiny, ``The rising trend of metaverse in education: Challenges, opportunities, and ethical considerations,'' \emph{PeerJ Computer Science}, vol.~9, p. e1252, 2023.

\bibitem{lepez2022metaverse}
C.~O. Lepez, ``Metaverse and education: A panoramic review,'' \emph{Metaverse Basic and Applied Research}, vol.~1, pp. 2--2, 2022.

\bibitem{zhou2022metaverse}
B.~Zhou, C.~Xi, G.~Li, and B.~Yang, ``Metaverse application in power systems,'' \emph{Power Generation Technology}, vol.~43, no.~1, pp. 1--9, 2022.

\bibitem{cheng2023metaverse}
S.~Cheng, ``Metaverse and social view,'' in \emph{Metaverse: Concept, Content and Context}.\hskip 1em plus 0.5em minus 0.4em\relax Springer, 2023, pp. 107--122.

\bibitem{falchuk2018social}
B.~Falchuk, S.~Loeb, and R.~Neff, ``The social metaverse: Battle for privacy,'' \emph{IEEE technology and society magazine}, vol.~37, no.~2, pp. 52--61, 2018.

\bibitem{kuru2023metaomnicity}
K.~Kuru, ``Metaomnicity: Toward immersive urban metaverse cyberspaces using smart city digital twins,'' \emph{IEEE Access}, vol.~11, pp. 43\,844--43\,868, 2023.

\bibitem{prakash2023security}
R.~Prakash, G.~R. Nayar, and T.~Thomas, ``Security risk assessment of metaverse based healthcare systems based on common vulnerabilities and exposures (cve),'' in \emph{2023 IEEE International Conference on Recent Advances in Systems Science and Engineering (RASSE)}.\hskip 1em plus 0.5em minus 0.4em\relax IEEE, 2023, pp. 1--10.

\bibitem{deng2023metaverse}
Y.~Deng, Z.~Weng, and T.~Zhang, ``Metaverse-driven remote management solution for scene-based energy storage power stations,'' \emph{Evolutionary Intelligence}, vol.~16, no.~5, pp. 1521--1532, 2023.

\bibitem{musamih2022nfts}
A.~Musamih, K.~Salah, R.~Jayaraman, I.~Yaqoob, D.~Puthal, and S.~Ellahham, ``Nfts in healthcare: vision, opportunities, and challenges,'' \emph{IEEE consumer electronics magazine}, vol.~12, no.~4, pp. 21--32, 2022.

\bibitem{wang2023nft}
M.~Wang and N.~Lau, ``Nft digital twins: a digitalization strategy to preserve and sustain miao silver craftsmanship in the metaverse era,'' \emph{Heritage}, vol.~6, no.~2, pp. 1921--1941, 2023.

\bibitem{nair2023unique}
V.~Nair, W.~Guo, J.~Mattern, R.~Wang, J.~F. O'Brien, L.~Rosenberg, and D.~Song, ``Unique identification of 50,000+ virtual reality users from head \& hand motion data,'' in \emph{32nd USENIX Security Symposium (USENIX Security 23)}, 2023, pp. 895--910.

\bibitem{freire2019deep}
D.~Freire-Obreg{\'o}n, F.~Narducci, S.~Barra, and M.~Castrillon-Santana, ``Deep learning for source camera identification on mobile devices,'' \emph{Pattern Recognition Letters}, vol. 126, pp. 86--91, 2019.

\bibitem{khoshgoftaar1994comparative}
T.~M. Khoshgoftaar, D.~L. Lanning, and A.~S. Pandya, ``A comparative study of pattern recognition techniques for quality evaluation of telecommunications software,'' \emph{IEEE Journal on Selected Areas in Communications}, vol.~12, no.~2, pp. 279--291, 1994.

\bibitem{zolkipli2011approach}
M.~F. Zolkipli and A.~Jantan, ``An approach for malware behavior identification and classification,'' in \emph{2011 3rd international conference on computer research and development}, vol.~1.\hskip 1em plus 0.5em minus 0.4em\relax IEEE, 2011, pp. 191--194.

\bibitem{varghese2022digital}
S.~A. Varghese, A.~D. Ghadim, A.~Balador, Z.~Alimadadi, and P.~Papadimitratos, ``Digital twin-based intrusion detection for industrial control systems,'' in \emph{2022 IEEE International Conference on Pervasive Computing and Communications Workshops and other Affiliated Events (PerCom Workshops)}.\hskip 1em plus 0.5em minus 0.4em\relax IEEE, 2022, pp. 611--617.

\bibitem{keren2016convolutional}
G.~Keren and B.~Schuller, ``Convolutional rnn: an enhanced model for extracting features from sequential data,'' in \emph{2016 International Joint Conference on Neural Networks (IJCNN)}.\hskip 1em plus 0.5em minus 0.4em\relax IEEE, 2016, pp. 3412--3419.

\bibitem{chung2015recurrent}
J.~Chung, K.~Kastner, L.~Dinh, K.~Goel, A.~C. Courville, and Y.~Bengio, ``A recurrent latent variable model for sequential data,'' \emph{Advances in neural information processing systems}, vol.~28, 2015.

\bibitem{hochreiter1998vanishing}
S.~Hochreiter, ``The vanishing gradient problem during learning recurrent neural nets and problem solutions,'' \emph{International Journal of Uncertainty, Fuzziness and Knowledge-Based Systems}, vol.~6, no.~02, pp. 107--116, 1998.

\bibitem{chung2014empirical}
J.~Chung, ``Empirical evaluation of gated recurrent neural networks on sequence modeling,'' \emph{arXiv preprint arXiv:1412.3555}, 2014.

\bibitem{dey2017gate}
R.~Dey and F.~M. Salem, ``Gate-variants of gated recurrent unit (gru) neural networks,'' in \emph{2017 IEEE 60th international midwest symposium on circuits and systems (MWSCAS)}.\hskip 1em plus 0.5em minus 0.4em\relax IEEE, 2017, pp. 1597--1600.

\bibitem{HeaterHel_Data}
P.~Akella, ``Digital twin gadget health,'' \url{https://www.kaggle.com/datasets/prasannaakella/digital-twin-gadget-health}, 2021.

\bibitem{fernando2005cloning}
O.~N.~N. Fernando, G.~Saito, U.~Duminduwardena, Y.~Tanno, and M.~Cohen, ``Cloning and teleporting avatars across workstations and mobile devices in collaborative virtual environments: Clipboard operations for virtual reality,'' \emph{Proc. ICIA}, vol.~5, 2005.

\end{thebibliography}

\end{document}